\documentclass[%
reprint,
 amsmath,amssymb,
 aps,
 twocolumn,
 prb,
]{revtex4-2}

\usepackage{graphicx}% Include figure files
\usepackage{dcolumn}% Align table columns on decimal point
\usepackage{bm}% bold math
\begin{document}

%\preprint{}
\preprint{APS/123-QED}

\title{Crossover of Superconductivity \\ 
across the antiferromagnetic end point \\
in FeSe$_{\rm 1-x}$S$_{\rm x}$ under pressure
}


\author{Kiyotaka Miyoshi,$^{1,2}$ Takanobu Nakatani,$^1$ Yumi Yamamoto,$^1$ Takumi Maeda,$^1$ Daichi Izuhara,$^1$
and Ikumi Matsushima,$^1$
}
%\thanks{}
%\affiliation{}
\affiliation{%
$^1$Department of Physics and Material Science, Shimane University, Matsue 690-8504, Japan
}%
\affiliation{%
$^2$Next Generation TATARA Co-Creation Center, Shimane University, Matsue 690-8504, Japan
}
%\address{%
%$^2$Next Generation TATARA Co-Creation Center, Shimane University, Matsue 690-8504, Japan
%}%


\date{\today}

\begin{abstract}
Temperature-pressure ($T$-$P$) phase diagrams of FeSe$_{\rm 1-x}$S$_{\rm x}$ were investigated 
by the measurements of dc magnetization ($M$) and electrical resistivity ($\rho$) under pressure, 
using single crystal specimens with $x$=0.04, 0.08 and 0.13. 
For all specimens, the $M$($T$) curves under pressure near the end point of the antiferromagnetic (AFM) phase 
are found to show a two-step diamagnetic response, which can be described as 
the sum of two diamagnetic components $M_1$($T$) and $M_2$($T$), 
indicating that two superconducting (SC) phases with different $T_{\rm c}$ values 
coexist within a pressure range of $\Delta$$P$$\sim$1 GPa. 
Moreover, the pressure dependence of the amplitudes of $M_1$($T$) and $M_2$($T$) indicates a continuous 
transfer of the volume fraction between the two SC phases. 
These behaviors suggest that a crossover of superconductivity occurs in conjunction with the emergence of AFM phase
and imply that the SC phases inside and outside the AFM phase could have different origins.

%These behaviors suggest that a crossover of superconductivity occurs in conjunction with the 
%emergence of AFM phase, and that the SC phases inside 
%and outside the AFM phase have different origins.
%to be expressed as sum of two components $M_1$($T$) and $M_2$($T$), each of which 
%exhibits a diamagnetic behavior below $T_{\rm c1}^{\rm dia}$ and $T_{\rm c2}^{\rm dia}$, respectively, 
%Moreover, the diamagnetic amplitude of $M_1$($T$) and $M_2$($T$) 
%We observed a crossover of the superconductivity with 
%increasing $P$ near the end point of the antiferromagnetic (AFM) phase, where 
%two superconducting phases coexist within a pressure width of $\Delta$$P$$\sim$1 GPa, 
%having different $T_{\rm c}$ values. 
%These results suggest that the superconducting phases inside 
%and outside the AFM phase have different origins. 
%Across the magnetic phase boundary,
%In the pressure range near the magnetic quantum critical point (QCP), 
%The obtained $T$-$P$ phase diagrams suggest that the superconducting phases inside 
%the crossover pressure range of
%two superconducting phases, through the systematic $M$($T$) measurements 
%under pressure. 
%it is found that a supercondcuting phase gradually disappears but another supercondcuting phase 
%appears with increasing pressure through the 
\end{abstract}

%\keywords{}

\maketitle

\section{\label{sec:level1}Introduction}
Since the discovery of superconductivity in LaFeAsO$_{\rm 1-x}$F$_{\rm x}$\cite{kamihara}, 
a wide variety of iron-based superconductors have been discovered. 
The emergence of unconventional superconductivity, driven by competition or cooperation 
with antiferromagnetic (AFM) and nematic phases, has attracted intensive research, 
aiming to uncover the pairing mechanism\cite{hosono,shibauchi1,fernandes}. 
Indeed, it is crucial to establish the phase diagrams of various iron-based superconductors, 
as fluctuations arising from the ordered phases adjacent to the superconducting (SC) 
phase are promising candidates for the pairing glue. 
In this context, FeSe is an important subject for exploring the $T$-$P$ 
phase diagram. This is because the nematic phase is decoupled from the AFM 
phase at ambient pressure unlike in other iron-based superconductors, where 
AFM order is intertwined with nematic correlation, such as in 
AFe$_2$As$_2$ (A=Sr, Ba, Ca, Eu)\cite{rotter,kasahara2,
tori,colo,matsubayashi,terashima1}, 
LaFeAsO$_{\rm 1-x}$F$_{\rm x}$\cite{luetkens}, 
NaFe$_{\rm 1-x}$A$_{\rm x}$As (A=Co\cite{parker,wang1}, Cu\cite{wang2}) 
and Sr$_2$VO$_3$FeAs\cite{ueshima,holenstein}. In addition, a fourfold enhancement 
is achieved in $T_{\rm c}$ under pressure\cite{masaki,medvedev,margadonna,braithwaite,
miyoshi09,miyoshi14}. 
FeSe has been revealed to have an intriguing $T$-$P$ phase diagram, 
where $T_{\rm c}$ increases rapidly after the nematic phase disappears at $\sim$2 GPa\cite{miyoshi14}, 
accompanied by the emergence of AFM phase above 1.2 GPa\cite{terashima2,sun,wang3,kothapalli,khasanov1}, 
indicating that the three phases compete with each other. 

Isovalently substituted FeSe$_{\rm 1-x}$S$_{\rm x}$ is a more desirable material 
to verify which fluctuation dominates the superconductivity 
or to search for nematic fluctuation-mediated superconductivity rather than FeSe, 
since the AFM phase appears at even higher pressure 
than the pressure of the nematic end point\cite{matsuura}, 
so that both phases are well separated.
In FeSe$_{\rm 1-x}$S$_{\rm x}$, whereas the nematic phase is suppressed by the S-substitution 
toward the nematic quantum critical point at $x_{\rm c}$$\sim$0.17\cite{licciardello}, 
$T_{\rm c}$ is found to be maximized at $x$$\sim$0.1\cite{ishida}, where spin fluctuation is 
also strongly enhanced\cite{wiecki1}. 
From various measurements and theoretical studies, the SC gap structure is 
thought to be of superconductivity mediated by spin fluctuation for pure FeSe ($x$=0)\cite{kang,sprau,befatto}, 
while significant changes in the electronic structure are observed across $x_{\rm c}$\cite{sato,hanaguri,coldea}, 
and nematic fluctuation-mediated pairing are suggested for $x$$>$$x_{\rm c}$\cite{nag}. 
Under pressure, the electronic structure has been investigated by the 
measurements of nuclear magnetic resonance\cite{kuwayama1,rana,kuwayama2} and quantum oscillation\cite{reiss}, 
suggesting the reconstitution of Fermi surfaces across the nematic end point\cite{yamakawa} and 
the distinction of two superconductivities under presence or absence of nematicity. 

For the further understanding of the superconductivity, 
it is highly desirable to study the evolution not only across the nematic end point 
but also across the end point of the AFM phase.
%to study the evolutions of superconductivity not only across the nematic end point 
%but also across the end point of the AFM phase is highly desired. 
For the purpose, microscopic measurements are necessary to be performed above 3 GPa, 
although experimental difficulties 
are faced especially in observing Fermi surface under high pressure due to the restricted 
experimental approaches at present. 
To investigate $T$-$P$ phase diagram of FeSe$_{\rm 1-x}$S$_{\rm x}$, 
electrical resistivity ($\rho$) measurements have been performed under 
pressure up to $\sim$8 GPa using a cubic anvil apparatus (CAA) which generates 
hydrostatic pressure\cite{matsuura}, 
in addition to the detailed $\rho$($T$) measurements below 2 GPa\cite{xiang}. 
Through these studies, a notable trend that the 
separation between the nematic and AFM phases becomes remarkable in the specimens with higher $x$ 
has been unveiled, but the details of the phase boundaries of SC 
and AFM phases remain unclear due to the large pressure intervals 
of $\sim$1 GPa in the measurements\cite{matsuura}. 

In the present work, we have performed the measurements of 
dc magnetization and electrical resistivity 
under pressure up to 6 GPa using single crystal specimens 
of FeSe$_{\rm 1-x}$S$_{\rm x}$ ($x$=0.04, 0.08 and 0.13)
to establish the $T$-$P$ phase diagram. 
We report the continuous evolution of the superconductivity across the end point of the AFM phase, 
as evidenced by the $M$($T$) curves, which exhibit a two-step diamagnetic response, suggesting that 
the superconducting phase gradually evolves to another one with a different $T_{\rm c}$ value. 
$T$-$P$ phase diagrams, where three superconducting phases SC1, SC2 and SC3 
appear, are proposed. 
\begin{figure*}[t]
\includegraphics[width=17cm]{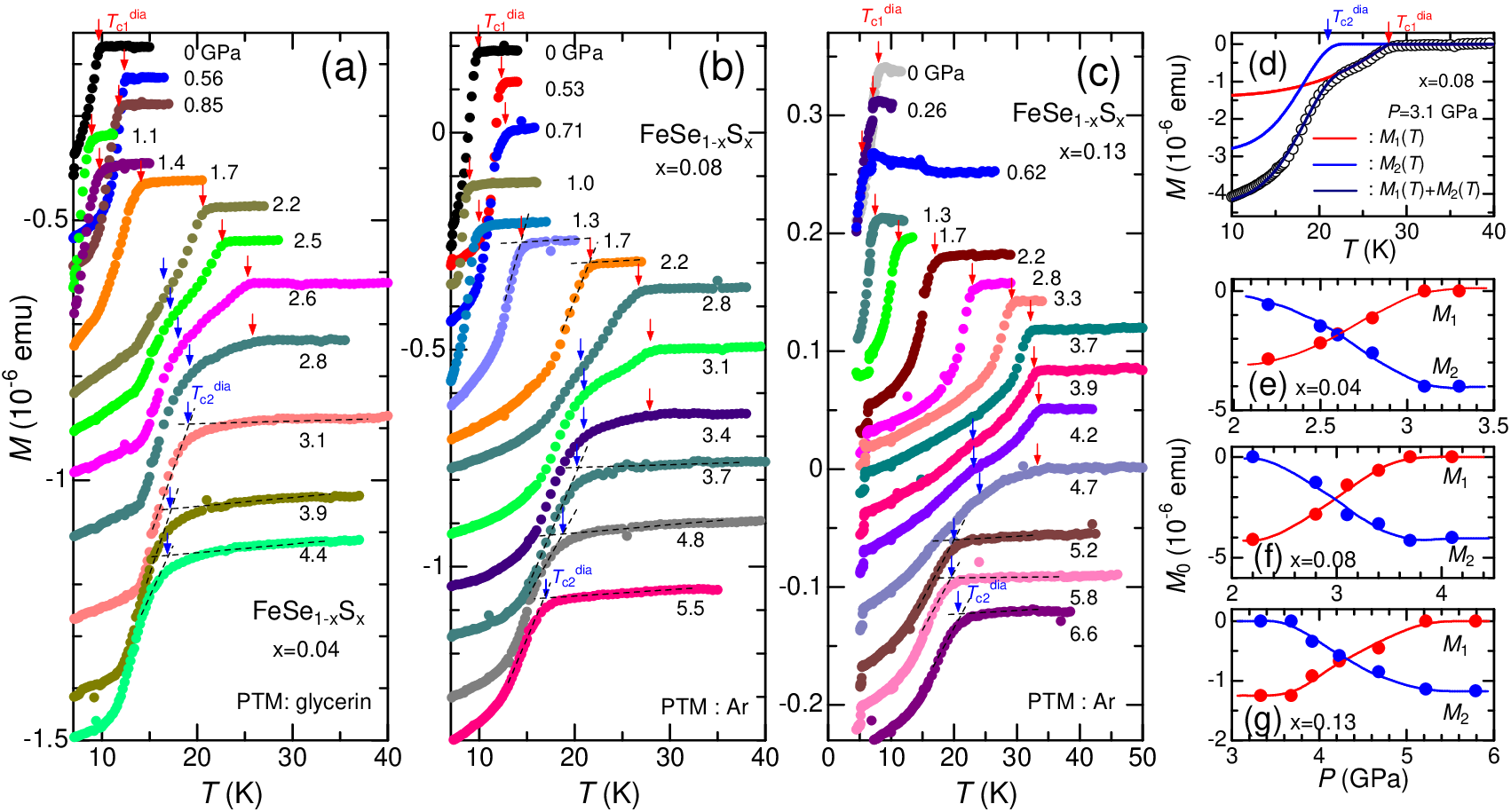}% Here is how to import EPS art
\caption{Temperature ($T$) dependence of zero-field-cooled dc magnetization ($M$) 
for FeSe$_{\rm 1-x}$S$_{\rm x}$ with $x$=0.04 (a), 0.08 (b) 
and 0.13 (c) measured by 
applying a magnetic field of 20 Oe at various pressures above 3-5 K. 
The data are intentionally shifted along the longitudinal 
axis for clarity. A jump in the $M$($T$) curves observed at low temperature for $x$=0.13 
is due to SC transition of a Pb manometer. 
(d) $M$($T$) curve for x=0.08 at $P$=3.1 GPa 
consisting of two components $M_{\rm 1}$($T$) (red solid line) and $M_{\rm 2}$($T$) (blue solid line), 
each of which shows a diamagnetic behavior below $T_{\rm c1}^{\rm dia}$ and $T_{\rm c2}^{\rm dia}$, respectively. 
We used Eq. (1) as a fitting function for $M$($T$) data. 
Plots of diamagnetic amplitude $M_0$ for $M_{\rm 1}$($T$) and $M_{\rm 2}$($T$) versus pressure for 
x=0.04 (e), 0.08 (f) and 0.13 (g). The solid lines are guide for the eyes. 
}
\end{figure*}

\section{\label{sec:level1}methods}
Single crystal specimens of FeSe$_{\rm 1-x}$S$_{\rm x}$ ($x$=0.04, 0.08 and 0.13) 
were obtained by a chemical vapor transport method 
in a similar way to those described in previous studies\cite{miyoshi20,miyoshi23}. 
Phase purity of the specimens was checked by X-ray diffraction (XRD) measurements, as shown in 
Figs. S1(a)-S1(d) in Supplemental Material\cite{suppl}. 
We show the lattice constants estimated by the measurements in Figs. S1(e)-S1(f), 
which were consistent with those in a literature\cite{matsuura}. 
S content $x$ in FeSe$_{\rm 1-x}$S$_{\rm x}$ single crystals used in the measurements were
estimated by energy-dispersive X-ray spectroscopy (EDX) measurements using an analytical scanning
electron microscope. To determine the S content $x$, the measurements were done at 
different 10 points and averaged values were adopted as $x$ values. 
The variation in $x$ at each measurement point was within 3-5$\%$.  
Magnetic measurements under high pressure were done by using a miniature
diamond anvil cell, which was combined with a sample rod of 
a commercial SQUID magnetometer. 
We used a CuBe gasket with a 0.3mm$\phi$ gasket hole, where a platelet FeSe$_{\rm 1-x}$S$_{\rm x}$ single
crystal was loaded parallel to the culet plane of the diamond
anvil together with a small piece of high-purity Pb to realize
the in-situ determination of pressure. The magnetization data
for FeSe$_{\rm 1-x}$S$_{\rm x}$ and Pb were obtained by subtracting the magnetic
contribution of the DAC measured in an empty run from the
total magnetization. In the measurements, a magnetic field of 20 Oe is 
applied to FeSe$_{\rm 1-x}$S$_{\rm x}$ single crystals perpendicular to 
the crystal surface. 
The measurements have been successfully applied 
to investigate pressure effects on 
superconductivity in our previous studies\cite{miyoshi08,miyoshi09,miyoshi13,miyoshi14,miyoshi21}. 
%The details of the measurements are given in the literature\cite{miyoshi21}. 
As the pressure transmitting media (PTM), we used Ar, which is known to be a hydrostatic PTM\cite{tateiwa}, 
for the specimens with $x$=0.08 and 0.13, but glycerine for $x$=0.04, which is hydrostatic below the 
solidification pressure ($\sim$5 GPa)\cite{tateiwa}. 
Electrical resistivity measurements under pressure were done 
by a standard 4-probe technique using an opposed-anvil cell to 
generate high pressure\cite{kitagawa}. For the pressure cell, we used a NiCrAr gasket with a sample 
hole of 2.2 mm$\phi$, where a FeSe$_{\rm 1-x}$S$_{\rm x}$ single crystal was 
set together with a high-purity Pb wire for 
the in-situ determination of pressure from the $T_{\rm c}$ shift. 
Also, glycerine was used as PTM for the measurements. 

\section{Results and Discussion}
\subsection{dc magnetization}
\begin{figure}[h]
\includegraphics[width=8cm]{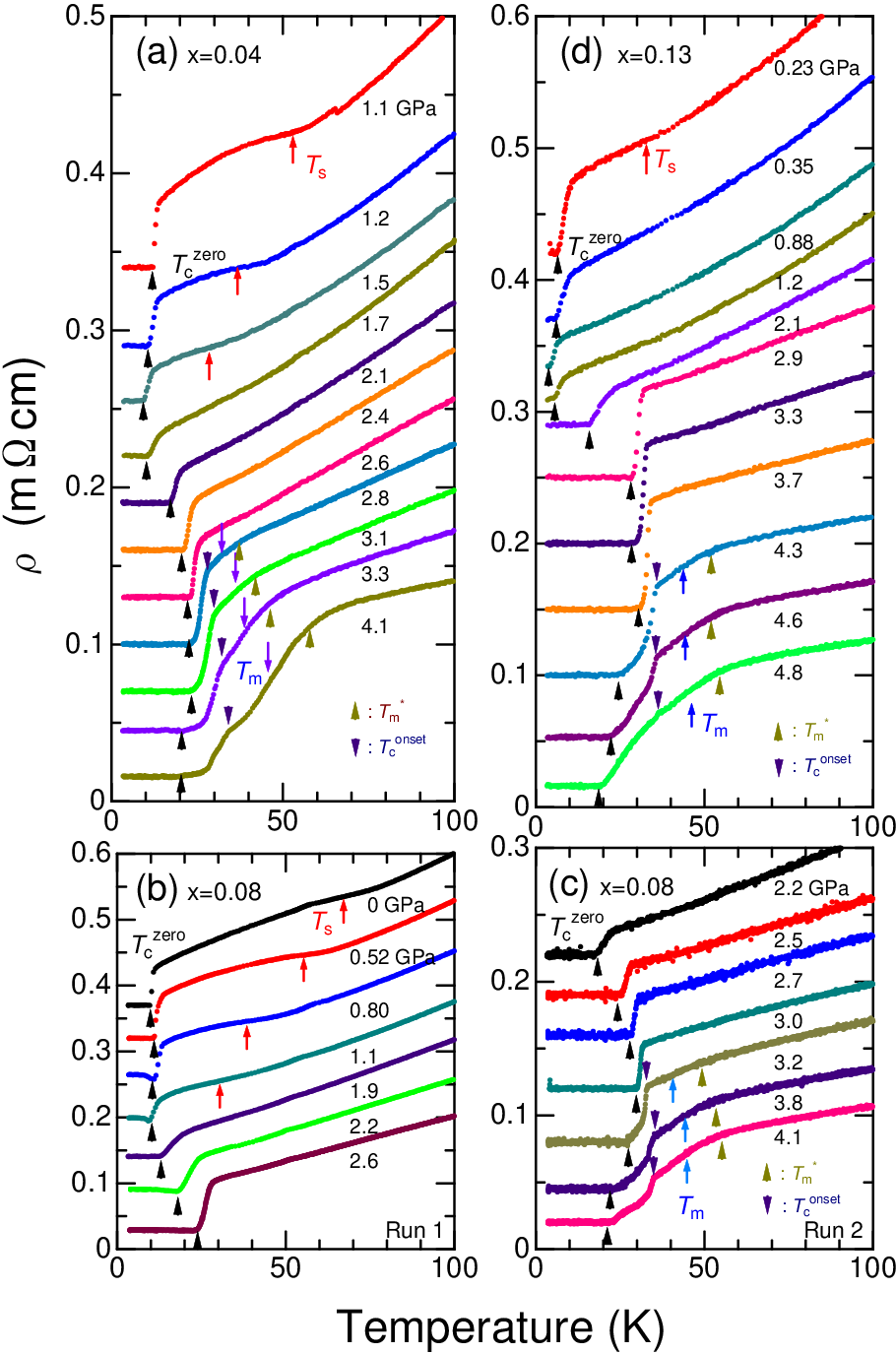}% Here is how to import EPS art
\caption{Temperature dependence of electrical resistivity $\rho$ for 
FeSe$_{\rm 1-x}$S$_{\rm x}$ with $x$=0.04 (a), 0.08 (run 1) (b), 0.08 (run 2) 
and 0.13 (c) measured at various pressures above $\sim$4 K using glycerin as the PTM. 
The data are intentionally shifted along the longitudinal 
axis for clarity. 
The black upward triangles indicate zero-resistive temperature $T_{\rm c}^{\rm zero}$. 
The red and blue arrows indicate the nematic ($T_{\rm s}$) and magnetic ($T_{\rm m}$) 
transition temperatures, respectively. 
}
\end{figure}
In Figs. 1(a)-1(c), we show zero-field-cooled dc magnetization versus temperature ($M$($T$)) 
data for $x$=0.04, 0.08 and 0.13 measured 
at various pressures. In Fig. 1(a), the $M$($T$) curve at ambient pressure shows a diamagnetic response 
below $\sim$10 K. The onset temperature of diamagnetic response, 
which is a reliable marker of $T_{\rm c}$ and we assign as $T_{\rm c1}^{\rm dia}$, 
shifts to a higher temperature at $P$=0.56 GPa 
but decreases down to $\sim$9 K at $P$=1.1 GPa, 
and then increases again above 1.1 GPa, showing a local maximum at $\sim$0.6 GPa. 
A local maximum below 1 GPa followed by a rapid increase in $T_{\rm c1}^{\rm dia}$ 
is a characteristic feature seen in FeSe ($x$=0)\cite{miyoshi14,miyoshi21} 
and also observed for $x$=0.08 in Fig. 1(b). 
In contrast, as seen in $M$($T$) curves for $x$=0.13 below 1 GPa in Fig. 1(c), 
$T_{\rm c1}^{\rm dia}$ ($\sim$ 8 K at ambient pressure) decreases with increasing pressure 
but turns to increase above 0.62 GPa, showing no local maximum. The behavior is consistent 
with that observed for $x$=0.12 in earlier studies\cite{kuwayama1,kuwayama2,xiang}. 

As seen in Figs. 1(a)-1(c), while $T_{\rm c1}^{\rm dia}$ shows a rapid 
increase above $\sim$1 GPa from the minimum value, we should note that a hump-like 
anomaly commonly begins to appear below $T_{\rm c1}^{\rm dia}$ 
in the $M$($T$) curves at $P$=2.2, 2.8 and 3.9 GPa for $x$=0.04, 0.08 and 0.13, respectively. 
In the figures, we assign the hump temperature as $T_{\rm c2}^{\rm dia}$, since 
the hump anomaly is thought to correspond to the onset of the second diamagnetic response 
in the $M$($T$) curves, in other words, the diamagnetic response 
occurs in two steps. The behavior indicates that 
there exist two distinct superconducting phases with different $T_{\rm c}$s, 
so that the $M$($T$) curve can be expressed as sum of $M_{\rm 1}$($T$) and $M_{\rm 2}$($T$), 
each of which exhibits a diamagnetic response below $T_{\rm c1}^{\rm dia}$ and $T_{\rm c2}^{\rm dia}$, 
respectively, as shown in Fig. 1(d) for $x$=0.08 at $P$=3.1 GPa as an example. 
$M_{\rm 1}$($T$) and $M_{\rm 2}$($T$) curves are individually described by the following 
phenomenological expression for $T$$\leq$$T_{\rm c}$, 
\begin{equation}
M(T)=-M_{0}\left\{1-\left(\frac{T}{T_{\rm c}}\right)^{n}\right\}^{m}, 
\end{equation}
%where $M_0$ corresponds to the diamagnetic amplitude at $T$=0 K and $T_{\rm c}$ is the 
%critical temperature. 
and $M$($T$)=0 for $T$$\geq$$T_{\rm c}$. 
This form has been employed, with a positive sign, as a flexible fitting function for the temperature evolution 
of magnetization in magnetic materials\cite{madhogaria,zivkovic}. 
In Figs. 1(a)-1(c), the coexistence of two superconducting phases 
can be seen in some $M$($T$) curves within a pressure range of $\sim$1 GPa, e.g., 
those at $P$=2.8, 3.1 and 3.4 GPa for $x$=0.08. 

We display $M$($T$) curves which are composed of two diamagnetic components 
in Figs. S2(a)-S2(i) in Supplemental Material\cite{suppl}, 
where $M$($T$) curves are expressed 
as sum of $M_{\rm 1}$($T$) and $M_{\rm 2}$($T$) in a similar manner shown in Fig. 1(d). 
It should be noted for $x$=0.04 in Figs. S2(a)-S2(d) that $M_{\rm 1}$($T$) which becomes diamagnetic 
below $T_{\rm c1}^{\rm dia}$ decreases the diamagnetic amplitude with increasing pressure, whereas 
$M_{\rm 2}$($T$) increases the amplitude with increasing pressure. 
Similar pressure evolutions of $M_{\rm 1}$($T$) and $M_{\rm 2}$($T$) are 
also seen for $x$=0.08 and 0.13. 
These features are confirmed in Fig. 1(e)-1(g), where 
pressure variations of $M_0$, i.e., diamagnetic amplitude at $T$=0 K, 
for $M_{\rm 1}$($T$) and $M_{\rm 2}$($T$) are plotted for $x$=0.04, 0.08 and 0.13. 
In the figures, the amplitude $M_0$ of $M_{\rm 1}$($T$) decreases, 
while that of $M_{\rm 2}$($T$) increases with increasing pressure, 
indicating a continuous evolution of the superconductivity from the superconductivity 
represented by $M_{\rm 1}$($T$) to that represented by $M_{\rm 2}$($T$). Figures 1(e)-1(g) demonstrate 
continuous transfers in the volume fraction of the superconductivity within a pressure range of $\sim$1 GPa. 

The observed two-step transition reminds us of the successive superconducting transition observed 
in ceramic superconductors, where superconducting order develops from intra- to 
intergrain region due to weak-link effects between grains\cite{kawachi,matsuuramoto}. 
Since the specimens used in the present study are single crystals, weak link effects could be realized 
between cleavable layers due to the layered structure. 
Then, one may consider that the two-step transition originates from weak link effects between 
layers, i.e., SC transition first occurs within layers below $T_{\rm c1}^{\rm dia}$ and then interlayer 
coupling occurs below $T_{\rm c2}^{\rm dia}$. 
However, since the magnetic field was applied perpendicular to the layers in the single crystal,
the diamagnetic response should not be significantly affected, 
even if interlayer superconducting coupling develops well below $T_{\rm c1}^{\rm dia}$. 
Moreover, a two-step SC transition induced by weak-link effects cannot account for 
a continuous transfer in the diamagnetic amplitudes with increasing pressure observed in the present study. 
Also, one may suspect that the two-steps superconducting 
transition arises from the sample inhomogeneity. However, this possibility can be ruled out 
by our XRD and EDX measurements as well as by the sharp SC transition observed in the specimens. 
In addition, specimens with inhomogeneous $x$ could exhibit two $T_{\rm c}$s but 
the volume fraction of each SC phase would remain unchanged with pressure. 
The continuous evolution of the superconductivity observed in the present study 
possibly originates from spatial phase separation near the 
end point of the AFM phase. In this case, another SC phase, which emerges in coexistence with the 
AFM phase characterized by a different $T_{\rm c}$, begins to appear below the pressure of the end point of the 
AFM phase, and then evolves while the original SC phase gradually disappears above the pressure. 
Finally, we note that the reason why we used zero-field-cooled magnetization rather than field-cooled magnetization 
for the analysis is that the magnitude of the field-cooled signal is very small due to 
the limited sample size in the DAC. We show field-cooled magnetization versus temperataure curve for 
$x$=0.08 under pressure of 3.1 GPa together with zero-field cooled magnetization data as an example in 
Fig. S3(a) in the Suppelmentary Material\cite{suppl}. 
%transfer of the volume fraction in the diamagnetic amplitude. 
%In this way, we rule out the above possibilities. 
%Finally, we note here the reason why we used zero-field-cooled magnetization, not field-cooled magnetization, 
%for the analysis, even though field-cooled magnetization is equilibrium magnetization reflecting Meissner 
%effect. This is because a diamagnetic signal of field-cooled magnetization from a small piece of 
%FeSe$_{\rm 1-x}$S$_{\rm x}$ single crystal in the sample space of the DAC we used is so small 
%to detect as shown in Fig. S3(a) in the Supplemental Material\cite{suppl}. 

\subsection{Electrical resistivity}
In our magnetic measurements, we observed that the SC phase continuously evolves 
into another one with a different $T_{\rm c}$ under pressure in all specimens. 
It is essential to clarify the relationship between the continuous evolution of superconductivity 
and the emergence of the AFM phase 
in the phase diagram. 
%It is important to confirm where the crossover occurs in the phase diagram. 
Considering the pressure evolution of AFM phase reported in a previous study 
on FeSe$_{\rm 1-x}$S$_{\rm x}$ with $x$=0.04, 0.08 and 0.12\cite{matsuura}, 
we note that the pressure at which the continuous evolution of superconductivity occurs is close to 
the onset pressure of the AFM phase. 
To clarify whether there is an intimate correlation 
between the superconducting crossover and the end point of the AFM phase 
or not, and to construct the phase diagrams, we performed the measurements of 
$\rho$($T$) under various pressures. The results are shown in Figs. 2(a)-2(d). 
In the figures, $\rho$($T$) curves show zero-resistivity below $T_{\rm c}^{\rm zero}$, indicating 
a SC transition of FeSe$_{\rm 1-x}$S$_{\rm x}$. 
Also, the nematic transition is visible as a kink in $\rho$($T$) curves and the transition temperature 
$T_{\rm S}$ is determined from the peak of $d$$\rho$$/$$dT$ curve. 
As examples, we show $\rho$($T$) and $d$$\rho$$/$$dT$ curves at ambient pressure 
in Figs. S4(a)-S4(c) in Supplemental Material\cite{suppl}. 
As seen in Fig. 2(a), the nematic transition for $x$=0.04 is suppressed by the application of pressure and 
disappears above 1.7 GPa. On the other hand, we note that the transition for $x$=0.13 is suppressed only by 
the pressure of 0.35 GPa. 
\begin{figure}[h]
\includegraphics[width=8cm]{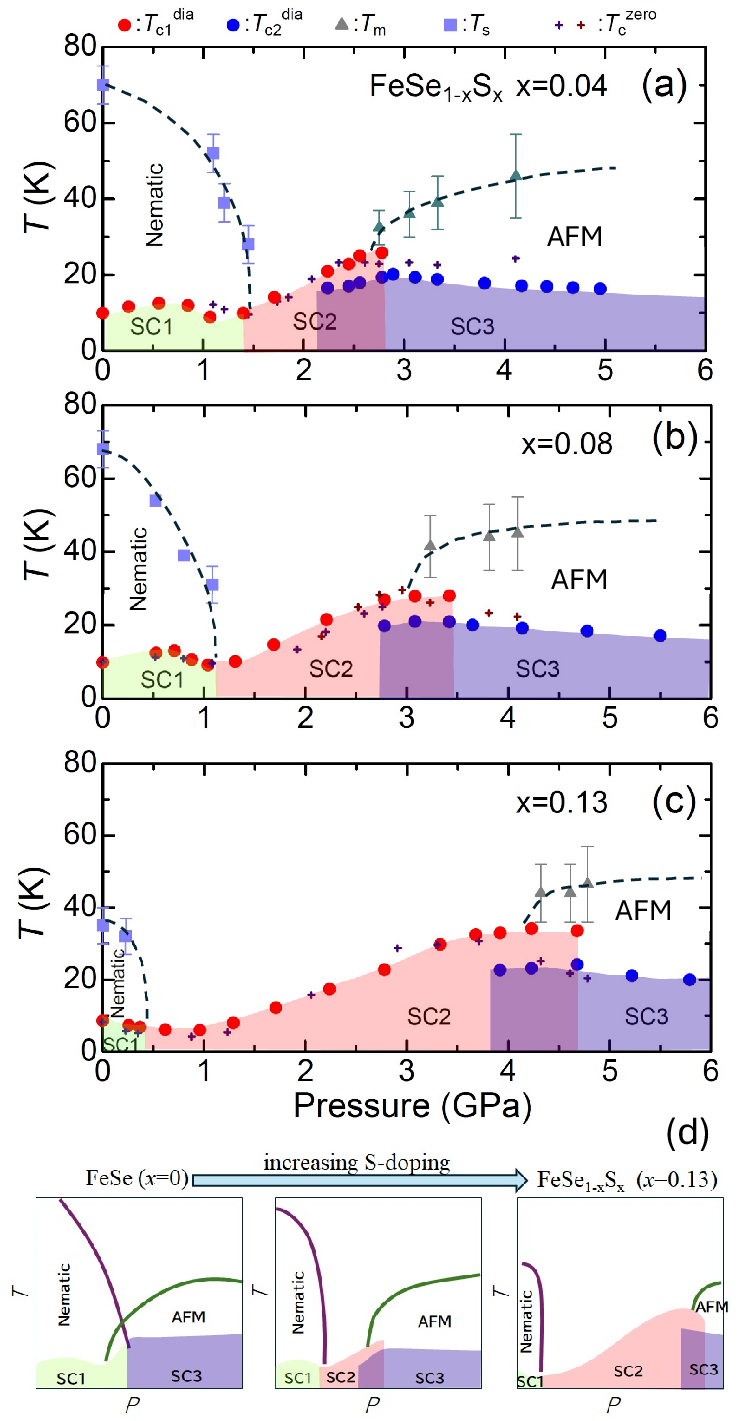}% Here is how to import EPS art
\caption{$T$-$P$ phase diagram of FeSe$_{\rm 1-x}$S$_{\rm x}$ with $x$=0.04 (a), 0.08 (b) 
and 0.13 (c). The broken lines are guide for the eyes. 
(d) Schematic view of the variation of the $T$-$P$ phase diagram of FeSe$_{\rm 1-x}$S$_{\rm x}$ 
with increasing $x$. 
}
\end{figure}

In a previous high-pressure study using a CAA to generate hydrostatic pressure, 
the transition temperature into the AFM phase, $T_{\rm m}$, was determined in most cases
from the peak in the $d$$\rho$$/$$dT$ curve \cite{matsuura}.
The peak was observed between the onset temperature 
of resistive drop associated with superconductivity $T_{\rm c}^{\rm onset}$ 
and $T_{\rm m}^{\rm *}$ below which $\rho$($T$) begins to decrease 
due to the evolution of AFM correlation which reduces the magnetic scattering. 
However, we failed to observe a clear peak of $d$$\rho$$/$$dT$ between $T_{\rm c}^{\rm onset}$ 
and $T_{\rm m}^{\rm *}$ in the present work, probably due to the difference in 
the degree of pressure homogeneity. 
Thus, we define a midpoint of $T_{\rm m}^{\rm *}$ and $T_{\rm c}^{\rm onset}$ as $T_{\rm m}$ 
in Figs. 2(a)-2(d). We estimated $T_{\rm c}^{\rm onset}$ ($T_{\rm m}^{\rm *}$) 
by extrapolating the initial slope of the 
$\rho$($T$) curve just below $T_{\rm c}^{\rm onset}$ ($T_{\rm m}^{\rm *}$)
to the $\rho$($T$) curve just above $T_{\rm c}^{\rm onset}$ ($T_{\rm m}^{\rm *}$) as shown in 
Figs. S5(a)-S5(c) in Supplemental Material\cite{suppl}. 
Here, we note that our magnetization measurements under pressure do not resolve the AFM transition, 
owing to the limited sample size, the restricted magnetic field, and the intrinsically 
weak magnetic anomaly at the AFM transition. We show the $M$($T$) curve for $x$=0.08 at $P$=3.7 GPa, 
including the high temperature region above 40 K where the AFM transition occurs 
in Fig. S3(b) in Supplemental Material\cite{suppl}. 

\subsection{$T$-$P$ phase diagram}
Figures 3(a)-3(c) display $T$-$P$ phase diagrams of FeSe$_{\rm 1-x}$S$_{\rm x}$, 
where the phase boundaries of SC phases are described by 
$T_{\rm c1}^{\rm dia}$ and $T_{\rm c2}^{\rm dia}$ obtained by the dc magnetization measurements. 
In the figures, it is found that the nematic phase shrinks with increasing $x$. 
The behavior agrees with that reported in an early work\cite{xiang}. 
Moreover, with increasing $x$, the AFM phase shifts to higher pressures, 
so that S substitution extends the pressure range where the SC phase exists 
alone in the intermediate region, without coexistence of nematic or AFM phases. 
%Moreover, the AFM phase moves to the higher pressure region with increasing $x$, 
%so that the S-substitution extends the region where the SC phase exists alone in the intermediate pressure range 
%without coexisting nematic or AFM phases. 
The evolution of AFM phase with increasing $x$ 
in the $T$-$P$ phase diagram and the pressures above which the AFM phase appear
are consistent with that observed by the measurements using a CAA\cite{matsuura}. 
We denote the SC phases below and above the nematic end point as SC1 and SC2, respectively, since two distinct 
superconductivities are expected to be realized due to the reconstitution of 
Fermi surfaces across the nematic end point\cite{kuwayama1,rana,kuwayama2,reiss,yamakawa}. 
It is also found that two SC phases with different $T_{\rm c}$s overlap within a 
pressure range of $\Delta$$P$$\sim$1 GPa, showing a crossover of the 
superconductivity. 
We also refer to the SC phase that appears after the crossover from SC2 as SC3. 
It should be noted that the occurrence of crossover from SC2 to SC3 
across the end point of AFM phase is confirmed in the phase diagrams, 
demonstrating that the mechanism of the superconductivity is 
different inside and outside the AFM phase. 
Although zero-resistivity is a good marker of SC transition, $T_{\rm c}^{\rm zero}$ is found to be between 
$T_{\rm c1}^{\rm dia}$ and $T_{\rm c2}^{\rm dia}$ in the crossover pressure region. 
A successive superconducting transition could not be detected by $\rho$($T$) measurements, 
because no further change can be observed in $\rho$($T$) once it reaches zero below the first 
$T_{\rm c}$. 
%We could not observe 
%a successive SC transition by $\rho$($T$) measurements, since nothing could be reflected on $\rho$($T$) 
%when $\rho$($T$) becomes zero below the first $T_{\rm c}$. 

\subsection{Discussion}
In Fig. 3(d), a schematic view of the evolution of the $T$-$P$ phase diagram 
with $x$ is shown. The SC2 phase is an intermediate phase between the SC1 and SC3 phases, 
which coexist with the nematic or AFM phases. 
Thus, for $x$=0, the intermediate SC2 phase is unlikely to exist, because the nematic and AFM phases 
are too close and overlap with each other, while the high-$T_{\rm c}$ SC phase is known to appear 
above the pressure at which the AFM phase terminates. 
One may consider that high-$T_{\rm c}$ superconductivity of the SC2 phase is 
enhanced by AFM and$/$or nematic fluctuations, since the SC2 phase is located between the nematic and magnetic phases. 
If this is the case, the $T_{\rm c}$ gap between the SC2 and the SC3 phases 
can be attributed to a suppression of AFM fluctuations within the AFM phase. 
We can see that the $T_{\rm c}$ gap becomes remarkable for $x$=0.13 in Fig. 3(c). 
This could be interpreted as evidence that AFM fluctuations become dominant 
in enhancing superconductivity as the nematic phase is suppressed for $x$=0.13. 
Nevertheless, the role of AFM fluctuations may remain unclear, 
since they are expected to be relatively weak. This expectation is based on 
the fact that the magnetic transition at $T_{\rm m}$ 
is of first-order nature due to spin-lattice coupling\cite{wang3,kothapalli}. 
It has also been reported that strong AFM fluctuations are present in the low-$T_{\rm c}$
SC1 dome below 1 GPa, whereas the high-$T_{\rm c}$
SC2 dome develops above 1 GPa with only weak AFM fluctuations\cite{kuwayama1,kuwayama2}.
%This could be viewed as evidence that the AFM fluctuations become dominant 
%to enhance the superconductivity due to the shrink of the nematic phase for $x$=0.13. 
%Nevertheless, the effect of AFM fluctuations may be uncertain, since 
%AFM fluctuations are expected to be relatively weak, considering the fact that 
%the transition at $T_{\rm m}$ is first-order type due to a spin-lattice coupling\cite{wang3,kothapalli}, as 
%also reported that the low-$T_{\rm c}$ SC1 dome below 1 GPa is accompanied by strong AFM fluctuations, whereas 
%high-$T_{\rm c}$ SC2 dome develops above 1 GPa with fairly weak AFM fluctuations\cite{kuwayama1,kuwayama2}. 

On the other hand, the effect of nematic fluctuations for the SC2 phase above the nematic end point 
is also unclear, since nematic fluctuations are thought to be quenched 
at the nematic end point probably due to the strong coupling to the lattice or local strain effects\cite{reiss}. 
Interestingly, in pristine FeSe, the collapse of nematic fluctuations above 1 GPa 
has been reported and it plays a marginal role for the high-$T_{\rm c}$ superconductivity\cite{massat}, 
stressing the difference from other iron-based superconductors, where both nematic and magnetic 
phases closely coexist and superconductivity is enhanced at the critical point. 
Recent microscopic measurements on the evolution of the precise gap structure with increasing $x$ across the nematic 
critical point $x_{\rm c}$ in FeSe$_{\rm 1-x}$S$_{\rm x}$ 
have revealed that the superconductivity for $x$$>$$x_{\rm c}$ is mediated by the nematic fluctuations\cite{nag}. 
To clarify the origin of SC2 and SC3, microscopic measurements under pressure 
which unveil the gap structure and 
Fermi surface reconstitution across the end point of AFM phase are essential as well. 
Occurrence of a Fermi surface reconstruction in the crossover region is likely in FeSe$_{\rm 1-x}$S$_{\rm x}$, 
as it was inferred from a sudden change in the Shubnikov-de Haas oscillation 
with the emergence of AFM phase in FeSe\cite{terashima3}.

Finally, we described the continuous evolution of the superconductivity across the 
end point of AFM phase as a crossover in this paper.
However, we note that it remains unclear whether this behavior represents a crossover 
or a phase transition, because we cannot exclude the possibility that the macroscopic 
volume fractions evolve gradually across a first-order transition accompanied by phase separation.
%However, we should note that 
%it is unclear whether this is a crossover or a phase transition, 
%since we cannot exclude the possibility of the gradual evolution of 
%macroscopic volume fractions across a first-order transition involving phase separation.

\section{Summary}
We have investigated $T$-$P$ phase diagram of FeSe$_{\rm 1-x}$S$_{\rm x}$ 
by the measurements of dc magnetization and electrical resistivity under pressure. 
It was found that $M$($T$) curves near the pressure where the AFM phase emerges 
show a diamagnetic response in two steps, 
indicating a coexistence of two distinct 
SC phases with different $T_{\rm c}$s. 
The characteristic behavior was observed for a pressure range of $\sim$1 GPa, where 
continuous transfer in the volume fraction of SC phases was indicated from the 
pressure variation of diamagnetic amplitude, suggesting a crossover of the superconductivity 
together with the emergence of AFM phase. 
Microscopic measurements which reveal the SC gap structure and the evolution of Fermi surface 
under pressure inside and outside the AFM phase are highly desired to elucidate the mechanism 
of superconductivities. 

\begin{acknowledgments}
The authors thank Shijo Nishigori, Takahiro Matsumoto, Kazuya Ando, Ryuichi Miyake, Kai Miyamoto
for technical assistance. This work was supported by technical assistance at Department of Materials Analysis, ICSR, 
Shimane University.
\end{acknowledgments}

\section*{data availability}
The data that support the findings of this article are openly available\cite{das}, embargo periods may apply.

%\begin{supplementalmaterial}
%See Supplemental Material at {\bf URLwillbeinsertedbypublisher} for additional figures. 
%\end{supplemental material}

\end{document}